\documentclass[conference]{IEEEtran}

\usepackage{myStyleIEEE}

\usepackage[margin=0.65in]{geometry}
\usepackage{booktabs}
\usepackage{tabularx}
\usepackage{amsmath}
\usepackage{threeparttable}
\usepackage{tablefootnote}

\DeclareSIUnit \voltampere { VA } %apparent power 
\DeclareSIUnit \var { var } %volt-ampere reactive - idle power 

\IEEEoverridecommandlockouts

\begin{document}

\title{Exploring Converter Control Duality in Microgrids: AC Grid-Forming vs DC Droop Control}

\renewcommand{\theenumi}{\alph{enumi}}

\newcommand{\jovan}[1]{\textcolor{magenta}{$\xrightarrow[]{\text{J}}$ #1}}
\newcommand{\ognjen}[1]{\textcolor{pPurple}{$\xrightarrow[]{\text{O}}$ #1}}

\author{\IEEEauthorblockN{Jovan~Krajacic\IEEEauthorrefmark{1}\IEEEauthorrefmark{2},
        Ognjen~Stanojev\IEEEauthorrefmark{1},
        Mario~Schweizer\IEEEauthorrefmark{1},
        Orcun~Karaca\IEEEauthorrefmark{1},
        Gabriela~Hug\IEEEauthorrefmark{2},
        Vladan Lazarević\IEEEauthorrefmark{1}}
        \IEEEauthorblockA{\IEEEauthorrefmark{1} ABB Corporate Research Center, Switzerland}
        \IEEEauthorblockA{\IEEEauthorrefmark{2} EEH - Power Systems Laboratory, ETH Zurich, Switzerland}
        \{jkrajacic, hug\}@eeh.ee.ethz.ch, \{ognjen.stanojev, mario.schweizer, orcun.karaca, vladan.lazarevic\}@ch.abb.com
\thanks{\textit{Corresponding Author:} Ognjen Stanojev, ognjen.stanojev@ch.abb.com}
}

\maketitle
\IEEEpeerreviewmaketitle

%ABSTRACT
\begin{abstract}
% 1. Introduce the control types
Power electronic converters are fundamental building blocks of both AC and DC microgrids, enabling the integration of renewable energy sources, energy storage systems, electronic loads, and electric vehicles.
%
% 1a. AC GFM
In AC microgrids, grid-forming control has emerged as a key concept for forming voltage and frequency in the grid.
%
% 1b. DC Droop
In contrast, converter control in DC microgrids has developed along the path of droop control, which is widely adopted for decentralized DC-bus voltage regulation and power sharing.
%
% 2. Research gap: the duality between them is unexplored
Although these control strategies share certain characteristics, their similarities remain largely unexplored due to the distinct physical domains in which they operate.
%
% 3. Addressing the research gap 
To bridge this gap, we introduce a novel perspective based on the concept of duality to reveal the underlying isomorphism between the two control approaches.
%
% 4. Paper contributions
We show that AC grid-forming and DC I--V droop control are duals of each other in several aspects, including: (i) the small-signal model of the converter; (ii) the inner current control structure; (iii) power-sharing mechanisms based on the AC swing equation and DC capacitor power balance; and (iv) disturbance signals and dynamic response. Theoretical analysis, validated through simulations on simple converter setups, illustrates these dualities and provides new insights towards a unified control design.
\end{abstract}

%INDEX TERMS
\begin{IEEEkeywords}
DC microgrids, AC microgrids, grid-forming control, droop control
\end{IEEEkeywords}

\section{Introduction} \label{sec:intro}
% 1. Introduce AC and DC microgrids
Driven by the imperative to adopt zero-emission energy resources, microgrids have emerged as an effective solution for integrating distributed energy resources into the energy system~\cite{5546958}. AC microgrids, in particular, have become a natural path for the legacy AC utility grid to transition from a centralized to a distributed structure~\cite{10942604}. Meanwhile, DC microgrids are experiencing rapid growth due to the increase in DC sources and loads such as photovoltaic panels, battery energy storage systems, and electric vehicles~\cite{7469404}. Nevertheless, AC microgrids remain dominant in both research and practice~\cite{JUSTO2013387}, while DC microgrids are attracting growing attention by eliminating the need for reactive power management and reducing the number of required AC/DC conversion stages, thus improving overall efficiency~\cite{su2024plug, 7152030}.

% 2. Control of AC and DC microgrids
The control of voltage source converters in AC systems has progressed from the well-established grid-following paradigm toward grid-forming (GFM) control, driven by the increasing penetration of power electronic interfaces and the corresponding reduction of synchronous machines~\cite{9714816}. 
In parallel, decentralized power sharing among source converters in DC microgrids has evolved along the path of droop control, enabling coordination without relying on communication between units~\cite{dcindustrie2_2024}. Although GFM terminology is still largely absent in DC microgrids, recent findings indicate that DC droop methods can exhibit GFM behavior over certain frequency ranges~\cite{forming2026}.
In this context, two main droop-control variants exist: voltage–current (V--I) droop and current–voltage (I--V) droop, where the former derives the voltage reference from the measured output current, whereas the latter computes the current reference based on the measured output voltage. 

Despite their distinct application domains and control complexities, AC GFM and DC droop control schemes share several fundamental characteristics, including inertia emulation, standalone operation, and voltage regulation~\cite{8894843}. The most widely studied approaches for providing these functionalities in AC systems include droop control~\cite{195899}, power synchronization control~\cite{PSC2010}, and virtual synchronous generator control~\cite{VSM2015}, while in DC systems, V--I droop extends to virtual DC machine control~\cite{Tan2016VDCM}, and I--V droop employs a virtual capacitance~\cite{8096714}.

% 3. Research gap: duality remains unexplored and obscured
Although similarities in droop control principles exist between DC I--V droop and AC GFM control, these strategies have traditionally evolved independently within their respective domains. As a result, the intrinsic duality between AC and DC microgrid control, particularly in terms of power-sharing dynamics, decentralized coordination and control design, has remained largely unexplored. While prior literature~\cite{8096714} has identified analogies between their power sharing mechanisms, i.e., the AC swing equation and DC capacitor power balance, a systematic verification of their dynamic equivalence is missing. 

% 4. Contributions
This paper establishes a unified framework for analyzing DC I--V droop and AC GFM control, placing them within a common theoretical foundation that reveals their underlying structural duality.
Motivated by recent duality-based approaches in the literature~\cite{9714816, 10224064}, the contributions of this work are threefold.
First, the converter interfaces are analyzed, demonstrating that they can be modeled equivalently under small-signal assumptions~\cite{658735, Liao2020}.
Second, the control loops are examined to establish the isomorphism between their inner current control structures and power-sharing mechanisms, and to identify the corresponding dual parameters and disturbances. 
Finally, the theoretical findings are validated through detailed simulations on realistic converter setups. By establishing this duality framework, this work provides a foundation for unified control design principles that bridge AC and DC  paradigms.

% 5. The remainder of the paper
The remainder of this paper is organized as follows. Section~\ref{sec:preliminaries} presents the small-signal model of the converters and their unified analysis. Section~\ref{sec:AC_GFM} introduces AC GFM control, and Section~\ref{sec:DC_IV} its DC counterpart, namely I--V droop control. The duality between these control strategies is established in Section~\ref{sec:duality} and validated through time-domain simulations in Section~\ref{sec:simulations}. Finally, Section~\ref{sec:concl} concludes the paper.

%%%%%%%%%%%%%%%%%%%%%%%%%%%%%%%%%%%%%%%%%%%%%%%%%%%%%%%%%%%%%%%%%%%%%%%%%%%
\begin{figure*}[t!]
    \centering
    \includegraphics[width=0.995\textwidth]{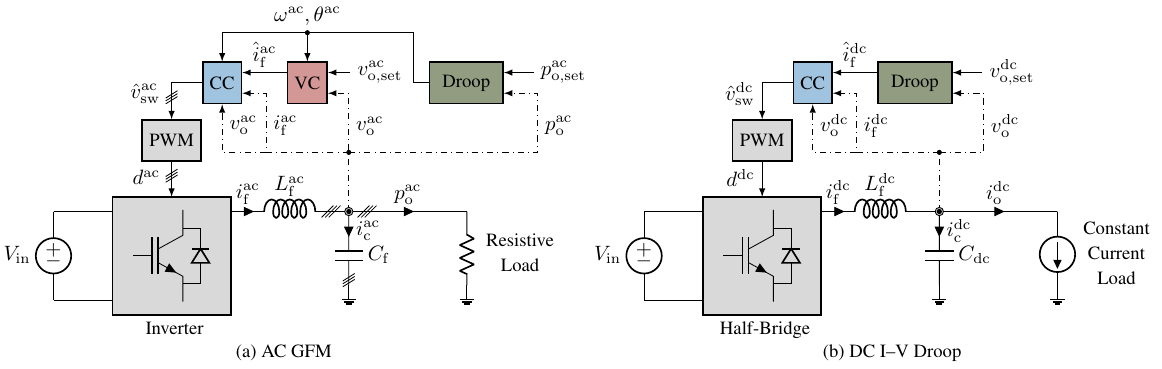}
    \vspace{-0.70cm}
    \caption{Source converters and their control structures connected to separate equivalent loads for (a) AC grid-forming (GFM) control with a resistive load and (b) DC I--V droop control with a constant current load, including the voltage controller (VC), current controller (CC), and pulse-width modulator (PWM).}
    \label{fig:converters}
    \vspace{-0.30cm}
\end{figure*}
%%%%%%%%%%%%%%%%%%%%%%%%%%%%%%%%%%%%%%%%%%%%%%%%%%%%%%%%%%%%%%%%%

\section{System Description and Modeling} \label{sec:preliminaries}
\subsection{AC and DC Converter Systems}
In this section, we present the considered AC and DC converter systems, as depicted in Fig.~\ref{fig:converters}. 
In both cases, the DC input voltage $V_\mathrm{in}$ is converted into a controllable output: a three-phase AC voltage $v_\mathrm{o}^\mathrm{ac}$ generated by a voltage source inverter in the AC case, and a scaled DC voltage $v_\mathrm{o}^\mathrm{dc}$ produced by a single half-bridge converter in the DC case, resembling a buck-type topology. Structurally, the half-bridge converter is equivalent to a single leg of a voltage source inverter.
The output voltage is regulated by a voltage controller (highlighted in red) in the AC case and by a droop controller (highlighted in green) in the DC case. In both systems, the inductor currents $i_\mathrm{f}^\mathrm{ac}$ and $i_\mathrm{f}^\mathrm{dc}$ are regulated by current controllers (highlighted in blue).
The pulse-width modulator (PWM) used in both converter models maps the reference voltages $\hat{v}_\mathrm{sw}^\mathrm{ac}$ and $\hat{v}_\mathrm{sw}^\mathrm{dc}$ to the corresponding duty-cycle signals $d^\mathrm{ac}$ and $d^\mathrm{dc}$, which are used to control the three-phase inverter in the AC case and the half-bridge converter in the DC case, respectively.

To facilitate a direct comparison between the three-phase AC converter and its single-phase DC counterpart, the AC converter is modeled in the synchronously rotating $dq$ reference frame. As a result, the control problem becomes two-dimensional and, after standard $dq$ decoupling is applied~\cite{658735}, the $d$- and $q$-axis dynamics become independent and symmetric. Consequently, the AC converter admits an equivalent single-input single-output representation per axis~\cite{Liao2020}. 
In the following analysis, we focus solely on the $d$-axis of the AC GFM converter, owing to its intrinsic duality with the DC I--V droop control algorithm, as shown in later sections. Due to the symmetry of the decoupled $dq$ dynamics, identical controller parameters apply to the $q$-axis. For brevity, the $d$-axis subscript is omitted in the subsequent AC expressions and figures.

\subsection{Small-Signal Modeling} \label{sec:preliminaries_ss}
\begin{figure*}[t!]
    \centering
    \includegraphics[width=0.995\textwidth]{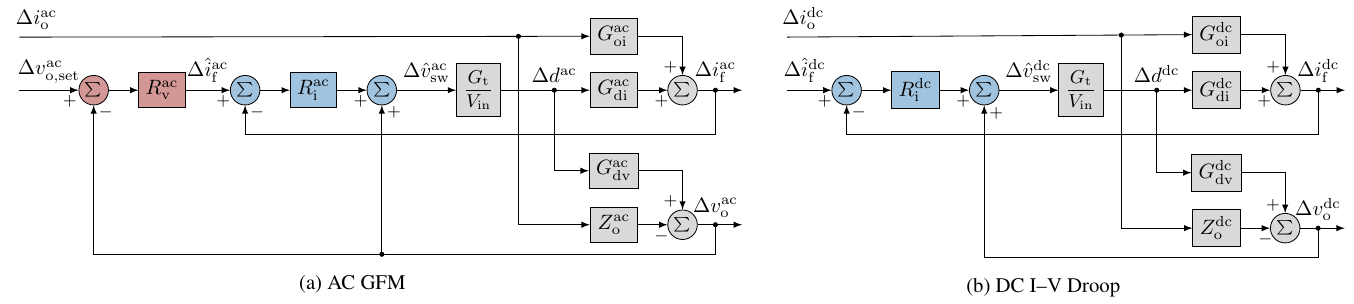}
    \vspace{-0.70cm}
    \caption{Small-signal control block diagrams of the source converters from Fig.~\ref{fig:converters}, including converter and PWM dynamics (gray), inner current (blue), and outer voltage (red) control loops for (a) AC GFM control and (b) DC I--V droop control. The PWM is modeled as a first-order delay $G_\mathrm{t}$ with a gain of $1/V_\mathrm{in}$.}
    \label{fig:control_loops}
    \vspace{-0.35cm}
\end{figure*}

The dynamics of both converters are inherently nonlinear due to their switching nature. To facilitate controller design and comparison, the systems are averaged over one switching period and linearized around a predefined operating point using small-signal perturbations~\cite{Erickson2001}. The resulting linearized equations describe the local dynamic behavior of the converters in the absence of control loops and can be transformed into the Laplace domain using the complex frequency variable $s = j\omega$.
Assuming a constant input voltage $V_\mathrm{in}$, the small-signal variations of the inductor current and terminal voltage for both the DC converter and the $d$-axis of the AC converter can be expressed in terms of the duty-cycle and output current perturbations as
\begin{align}
    \Delta i_{\mathrm{f}} &= G_\mathrm{di}(s)\Delta d + G_\mathrm{oi}(s)\Delta i_\mathrm{o}, \label{eq:ilf_general} \\
    \Delta v_\mathrm{o} &= G_\mathrm{dv}(s)\Delta d - Z_\mathrm{o}(s)\Delta i_\mathrm{o}, \label{eq:vdc_general}
\end{align}
where the operator $\Delta(\cdot)$ denotes small-signal perturbations around steady-state values. Here, the superscripts $(\cdot)^{\mathrm{ac}}_d$ and $(\cdot)^{\mathrm{dc}}$ are omitted for notational simplicity, since the derived expressions apply to both converter types. These relationships and the associated transfer functions are included as gray blocks in the small-signal models of the converters in Fig.~\ref{fig:control_loops}. For brevity, the explicit dependence on $s$ is omitted hereafter. The transfer functions in \eqref{eq:ilf_general} and \eqref{eq:vdc_general} are given by \cite{Liao2020}:
\begin{align}
	G_\mathrm{di} &= \frac{V_\mathrm{in}Y_{\mathrm{C}}}{1+Y_{\mathrm{C}}Z_{\mathrm{L}}},  \label{eq:Gdi_Goi}
	&G_\mathrm{oi} = \frac{1}{1+Y_{\mathrm{C}}Z_{\mathrm{L}}}, \\ 
	G_\mathrm{dv} &= \frac{V_\mathrm{in}}{1+Y_{\mathrm{C}}Z_{\mathrm{L}}}, 
	&Z_\mathrm{o} = \frac{Z_{\mathrm{L}}}{1+Y_{\mathrm{C}}Z_{\mathrm{L}}}, \label{eq:Gdv_Zo}
\end{align}
where $Z_{\mathrm{L}}$ represents the filter impedance, given by $sL_\mathrm{f}^\mathrm{ac}$ for the AC and $sL_\mathrm{f}^\mathrm{dc}$ for the DC converter, while $Y_{\mathrm{C}}$ denotes the capacitor admittance, equal to $sC_\mathrm{f}$ and $sC_\mathrm{dc}$, respectively.

Although the two converters exhibit equivalent open-loop behavior, their closed-loop dynamics differ due to the structure of their control loops, as illustrated in Figs.~\ref{fig:converters}~and~\ref{fig:control_loops}, which at first glance hinders a unified representation. Specifically, the AC GFM converter employs a dual-loop control structure, whereas the DC I--V droop converter adopts a single-loop structure. Additionally, the droop controllers serve different roles in the two systems: in the AC case, the droop controller regulates the emulated frequency $\omega^\mathrm{ac}$, whereas in the DC case, it generates the reference for the current controller $\hat{i}_\mathrm{f}^\mathrm{dc}$. Despite these differences, their underlying power-sharing mechanisms are fundamentally equivalent, as demonstrated in the following sections through a detailed analysis of their control structures.

%%%%%%%%%%%%%%%%%%%%%%%%%%%%%%%%%%%%%%%%%%%%%%%%%%%%%%%%%%%%%%%%%%%%%%%%%%%
%%%%%%%%%%%%%%%%%%%%%%%%%%%%%%%%%%%%%%%%%%%%%%%%%%%%%%%%%%%%%%%%%

\section{AC Grid-Forming Control} \label{sec:AC_GFM}
The AC GFM control analyzed in this work is implemented using a dual-loop architecture comprising an outer voltage loop and an inner current loop~\cite{9361257}, as illustrated in Fig.~\ref{fig:converters}a. The inner PI current controller is responsible for tracking the current reference $\hat{i}_\mathrm{f}^\mathrm{ac}$, while the outer PI voltage controller regulates the converter output voltage to the desired setpoint $v_\mathrm{o,set}^\mathrm{ac}$.
The droop controller is implemented in the form of active power–frequency droop for the $d$-axis, whereas the $q$-axis regulates the reactive power to maintain a constant terminal voltage magnitude. % Together, the $d$- and $q$-axes determine the setpoint voltage $v_\mathrm{o,set}^\mathrm{ac}$.
Under the $dq$ decoupling introduced in the previous section, active and reactive power are regulated independently. Hence, the following analysis focuses solely on the $d$-axis dynamics and control. Fig.~\ref{fig:control_loops}a presents the resulting small-signal control structure.

\subsection{Inner Current Control} \label{sec:cc_AC}
To tune the PI current controller $R_\mathrm{i}^\mathrm{ac}$ depicted in blue in Fig.~\ref{fig:control_loops}a, we derive the current-loop gain transfer function $T_\mathrm{i}^\mathrm{ac}$ following the approach in~\cite{Liao2020}. The function is obtained by opening the current feedback loop at the summing junction of the current controller and deriving the transfer from the current error signal $\Delta \hat{i}_\mathrm{f}^\mathrm{ac} - \Delta i_\mathrm{f}^\mathrm{ac}$ to the controlled variable $\Delta i_\mathrm{f}^\mathrm{ac}$:
\begin{equation}
	T_\mathrm{i}^\mathrm{ac}
	=
	\frac{G_\mathrm{di}^\mathrm{ac}G_\mathrm{t}/V_\mathrm{in}}
	     {1 - G_\mathrm{dv}^\mathrm{ac}G_\mathrm{t}/V_\mathrm{in}}
	R_\mathrm{i}^\mathrm{ac}
	=
	\frac{sC_\mathrm{f}G_\mathrm{t}}
	     {1 - G_\mathrm{t} + s^2L_\mathrm{f}^\mathrm{ac}C_\mathrm{f}}
	R_\mathrm{i}^\mathrm{ac}. \label{eq:Ti_AC}
\end{equation}

Following the internal model control principle~\cite{658735}, the inner current controller is designed such that the closed-loop current dynamics approximate unity gain over the bandwidth of the current controller. Within this frequency range, the effect of the output capacitance becomes negligible, and the current-loop dynamics are primarily governed by the inductor, provided that the $L_\mathrm{f}^\mathrm{ac}C_\mathrm{f}$ resonance lies beyond the current control bandwidth. Furthermore, the PWM dynamics $G_\mathrm{t}$, modeled as a first-order delay, can be approximated by unity in this frequency range. Assuming ideal feedforward of the output voltage $\Delta v_\mathrm{o}^\mathrm{ac}$, the current-loop gain can be approximated as a first-order system:
\begin{equation}
	T_\mathrm{i}^\mathrm{ac} \approx \frac{1}{sL_\mathrm{f}^\mathrm{ac}}\, R_\mathrm{i}^\mathrm{ac}. \label{eq:Tis_AC}
\end{equation}

Thus, internal model control yields the following parameters:
\begin{equation}
	R_\mathrm{i}^\mathrm{ac}
	=
	k_\mathrm{pi}^\mathrm{ac}\frac{1+sT_\mathrm{ii}^\mathrm{ac}}{sT_\mathrm{ii}^\mathrm{ac}},
	\quad \,\,
	k_\mathrm{pi}^\mathrm{ac} = \omega_{\beta_\mathrm{i}}^\mathrm{ac} L_\mathrm{f}^\mathrm{ac},
	\quad \,\,
	T_\mathrm{ii}^\mathrm{ac} = \frac{c_\mathrm{i}^\mathrm{ac}}{\omega_{\beta_\mathrm{i}}^\mathrm{ac}}, \label{eq:cc_ac} % 20
\end{equation}
where the current controller is parameterized by the desired current-loop bandwidth $\omega_{\beta_\mathrm{i}}^\mathrm{ac}$ and the integral time constant tuning factor $c_\mathrm{i}^\mathrm{ac}$.

\subsection{Outer Voltage Control} \label{sec:vc_AC}
After tuning the current controller, the $d$-axis control-loop analysis of the AC GFM converter proceeds with the PI voltage controller $R_\mathrm{v}^\mathrm{ac}$. Under the assumption that the current loop exhibits unity gain within the voltage controller bandwidth, the inductor current follows its reference, i.e., $\Delta i_\mathrm{f}^\mathrm{ac}=\Delta \hat{i}_\mathrm{f}^\mathrm{ac}$. Thus, the voltage dynamics are governed by the output capacitor dynamics of Fig.~\ref{fig:converters}a, which in the Laplace domain can be written as:
\begin{equation}
    \Delta i_\mathrm{c}^\mathrm{ac} = s C_\mathrm{f} \Delta v_\mathrm{o}^\mathrm{ac} = \Delta \hat{i}_\mathrm{f}^\mathrm{ac} - \Delta i_\mathrm{o}^\mathrm{ac}.
    \label{eq:AC_capacitor}
\end{equation}
Combining this with the voltage control law from Fig.~\ref{fig:control_loops}a,
\begin{equation}
    \Delta \hat{i}_\mathrm{f}^\mathrm{ac} =
    R_\mathrm{v}^\mathrm{ac}
    \left(\Delta v_\mathrm{o,set}^\mathrm{ac}-\Delta v_\mathrm{o}^\mathrm{ac}\right),
\end{equation}
yields the simplified voltage-loop gain $T_\mathrm{v}^\mathrm{ac}$, defined as the open-loop transfer function from the voltage error signal $\Delta v_\mathrm{o,set}^\mathrm{ac}-\Delta v_\mathrm{o}^\mathrm{ac}$ to the output voltage $\Delta v_\mathrm{o}^\mathrm{ac}$:
\begin{equation}
	T_\mathrm{v}^\mathrm{ac} \approx \frac{1}{sC_\mathrm{f}}\, R_\mathrm{v}^\mathrm{ac}.
\end{equation}

Following again the internal model control principle, the voltage controller is selected as:
\begin{equation}
	R_\mathrm{v}^\mathrm{ac}
	=
	k_\mathrm{pv}^\mathrm{ac}\frac{1+sT_\mathrm{iv}}{sT_\mathrm{iv}},
	\;\;  \quad
	k_\mathrm{pv}^\mathrm{ac} = \omega_{\beta_\mathrm{v}}^\mathrm{ac} C_\mathrm{f},
	\;\;  \quad
	T_\mathrm{iv}^\mathrm{ac} = \frac{c_\mathrm{v}^\mathrm{ac}}{\omega_{\beta_\mathrm{v}}^\mathrm{ac}}, \label{eq:vc_ac} % 2.5
\end{equation}
where the tuning is defined by the desired voltage-loop bandwidth $\omega_{\beta_\mathrm{v}}^\mathrm{ac}$ and the integral time constant tuning factor $c_\mathrm{v}^\mathrm{ac}$.

%%%%%%%%%%%%%%%%%%%%%%%%%%%%%%%%%%%%%%%%%%%%%%%%%%%%%%%%%%%%%%%%%%
\begin{figure}[b!]
    \centering
    \includegraphics[scale=0.9]{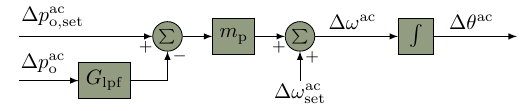}
    \vspace{-0.25cm}
    \caption{Small-signal droop control loop of AC GFM control.}
    \label{fig:droop_control_AC}
\end{figure}
%%%%%%%%%%%%%%%%%%%%%%%%%%%%%%%%%%%%%%%%%%%%%%%%%%%%%%%%%%%%%%%%%%

%%%%%%%%%%%%%%%%%%%%%%%%%%%%%%%%%%%%%%%%%%%%%%%%%%%%%%%%%%%%%%%%%%
\begin{figure*}[t!]
    \centering
    \includegraphics[width=0.995\textwidth]{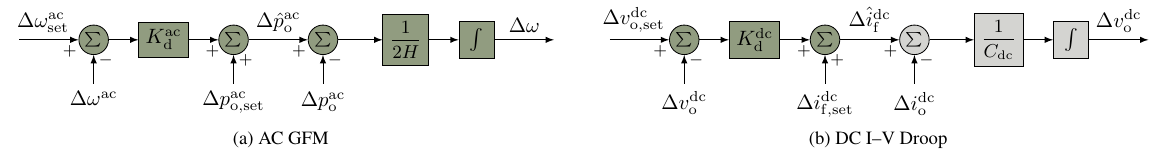}
    \vspace{-0.70cm}
    \caption{Dual power-sharing mechanisms of (a) AC GFM control and (b) DC I--V droop control.}
    \label{fig:droop_control}
    \vspace{-0.30cm}
\end{figure*}
%%%%%%%%%%%%%%%%%%%%%%%%%%%%%%%%%%%%%%%%%%%%%%%%%%%%%%%%%%%%%%%%%%

\subsection{Active Power-Frequency Droop Control} \label{sec:droop_AC}
Active power-frequency droop, hereinafter referred to as $P$--$\,\omega$ droop, adjusts the active power output of converters by emulating the frequency behavior of synchronous generators, thereby enabling decentralized synchronization and power sharing without requiring communication. The small-signal block diagram of the frequency droop controller implemented in this work is shown in Fig.~\ref{fig:droop_control_AC}, and its droop function is given~by:
\begin{equation}
    \Delta \omega^\mathrm{ac} = \Delta \omega_\mathrm{set}^\mathrm{ac} + m_\mathrm{p} \left(\Delta p_\mathrm{o,set}^\mathrm{ac} -  G_\mathrm{lpf} \Delta {p}_\mathrm{o}^\mathrm{ac}\right), \label{eq:AC_Droop}
\end{equation}
where $\Delta \omega^\mathrm{ac}$ denotes the calculated frequency of the internal reference frame and consequently of the ideally applied voltage vector $v_\mathrm{o}^\mathrm{ac}$, and $m_\mathrm{p}$ is the active power droop gain. The quantities $\Delta p_\mathrm{o,set}^\mathrm{ac}$ and $G_\mathrm{lpf}\Delta p_\mathrm{o}^\mathrm{ac}$ denote the active power setpoint and the filtered active power measurement, respectively, where filtering is implemented using a first-order low-pass filter:
\begin{equation}
    G_\mathrm{lpf}=\frac{\omega_\mathrm{c}}{\omega_\mathrm{c}+s}, \label{eq:lpf}
\end{equation}
with corner frequency $\omega_\mathrm{c}$.

As shown in the literature~\cite{Equivalence_VSM_Droop}, under small-signal assumptions, the $P$--$\,\omega$ law \eqref{eq:AC_Droop} can emulate the dynamic behavior of a synchronous generator. Specifically, by including the low-pass filter transfer function \eqref{eq:lpf}, \eqref{eq:AC_Droop} can be rewritten as~\cite{8398752}:
\begin{equation}
\begin{aligned}
    \frac{1}{\omega_\mathrm{c}m_\mathrm{p}} s \Delta \omega^\mathrm{ac} & =  
    \Delta p_\mathrm{o,set}^\mathrm{ac} - \Delta p_\mathrm{o}^\mathrm{ac} 
    + \frac{1}{m_\mathrm{p}} (\Delta \omega_\mathrm{set}^\mathrm{ac} - \Delta \omega^\mathrm{ac}) \\
    &+ \underbrace{\frac{1}{\omega_\mathrm{c}m_\mathrm{p}} s \Delta \omega_\mathrm{set}^\mathrm{ac}}_{\sigma_\omega}
    + \underbrace{\frac{1}{\omega_\mathrm{c}} s \Delta p_\mathrm{o,set}^\mathrm{ac}}_{\sigma_p}
\end{aligned}
\end{equation}
Assuming $\omega^\mathrm{ac} \approx 1$~p.u. and $\sigma_\omega = \sigma_p = 0$, the equation can be expressed in a form analogous to the swing equation, which governs the output power dynamics of the GFM converter:
\begin{equation}
    2H \ddt{\Delta \omega^\mathrm{ac}}= \Delta p_\mathrm{o,set}^\mathrm{ac}- \Delta p_\mathrm{o}^\mathrm{ac}+ K_\mathrm{d}^\mathrm{ac}\left(\Delta \omega_\mathrm{set}^\mathrm{ac} -  \Delta \omega^\mathrm{ac}\right), \label{eq:AC_Swing}
\end{equation}
as illustrated in Fig.~\ref{fig:droop_control}a. In this formulation, the equivalent inertia constant $H$ and damping coefficient $K_\mathrm{d}^\mathrm{ac}$ are related to the parameters in \eqref{eq:AC_Droop} as:
\begin{equation}
    H=\dfrac{1}{2\omega_\mathrm{c}m_\mathrm{p}}, \qquad 
    K_\mathrm{d}^\mathrm{ac} = \dfrac{1}{m_\mathrm{p}}. \label{eq:swing_parameters}
\end{equation}

\section{DC I--V Droop Control} \label{sec:DC_IV}
In contrast to AC GFM control, DC I--V droop control is implemented using a single-loop structure~\cite{lazarevic_dc_microgrids}, as shown in Fig.~\ref{fig:converters}b. Here, only the inner current control loop is explicitly implemented to regulate the inductor current $i_\mathrm{f}^\mathrm{dc}$ to its reference value $\hat{i}_\mathrm{f}^\mathrm{dc}$, while voltage regulation is achieved implicitly through the droop characteristic, which maps output voltage deviations $v_\mathrm{o,set}^\mathrm{dc} - v_\mathrm{o}^\mathrm{dc}$ to the current reference $\hat{i}_\mathrm{f}^\mathrm{dc}$. Fig.~\ref{fig:control_loops}b shows the small-signal form of the converter from Fig.~\ref{fig:converters}b.

\subsection{Inner Current Control} \label{sec:cc_DC}
The PI current controller design for the DC case follows the same procedure as for the AC GFM converter, since both systems regulate DC quantities under the considered modeling approach. Specifically, the current loop gain $T_\mathrm{i}^\mathrm{dc}$ can be derived using the small-signal model of the DC converter from Fig.~\ref{fig:control_loops}b:
\begin{equation}
	T_\mathrm{i}^\mathrm{dc}
	=
	\frac{G_\mathrm{di}^\mathrm{dc}G_\mathrm{t}/V_\mathrm{in}}
	     {1 - G_\mathrm{dv}^\mathrm{dc}G_\mathrm{t}/V_\mathrm{in}}
	R_\mathrm{i}^\mathrm{dc}
	=
	\frac{sC_\mathrm{f}G_\mathrm{t}}
	     {1 - G_\mathrm{t} + s^2L_\mathrm{f}^\mathrm{dc}C_\mathrm{f}}
	R_\mathrm{i}^\mathrm{dc}. \label{eq:Ti_DC}
\end{equation}

Comparing \eqref{eq:Ti_AC} and \eqref{eq:Ti_DC}, the current loop gains of the AC and DC converters are identical due to their isomorphic small-signal models derived in Section~\ref{sec:preliminaries_ss}. Consequently, the simplified current loop gain follows from the AC expression~\eqref{eq:Tis_AC}:
\begin{equation}
	T_\mathrm{i}^\mathrm{dc} \approx \frac{1}{sL_\mathrm{f}^\mathrm{dc}}\, R_\mathrm{i}^\mathrm{dc}. \label{eq:Tis_DC}
\end{equation}

Thus, the PI current controller $R_\mathrm{i}^\mathrm{dc}$ of the DC converter is tuned according to the dual internal model control principle of its AC counterpart, as discussed in Section~\ref{sec:cc_AC}:
\begin{equation}
	R_\mathrm{i}^\mathrm{dc}
	=
	k_\mathrm{pi}^\mathrm{dc}\frac{1+sT_\mathrm{ii}^\mathrm{dc}}{sT_\mathrm{ii}^\mathrm{dc}},
	\quad \,\,
	k_\mathrm{pi}^\mathrm{dc} = \omega_{\beta_\mathrm{i}}^\mathrm{dc} L_\mathrm{f}^\mathrm{dc},
	\quad \,\,
	T_\mathrm{ii}^\mathrm{dc} = \frac{c_\mathrm{i}^\mathrm{dc}}{\omega_{\beta_\mathrm{i}}^\mathrm{dc}}. \label{eq:cc_dc} % 20
\end{equation}

\subsection{I--V Droop Control} \label{sec:droop_DC}
Droop curves maintain the power balance within a DC grid by measuring the DC bus voltage and adjusting the power consumption or generation of active devices accordingly. This voltage-based control approach is recommended by emerging standards for DC microgrids~\cite{dcindustrie2_2024} and facilitates the parallel operation of supply and storage units, enabling decentralized coordination that scales to large systems.
In the case of I--V droop control implemented in this work, the droop control law, highlighted in green in Fig.~\ref{fig:droop_control}b, takes the following form:
\begin{equation}
    \Delta \hat{i}_\mathrm{f}^\mathrm{dc} =  \Delta i_\mathrm{f,set}^\mathrm{dc} + K_\mathrm{d}^\mathrm{dc} \left(\Delta v_\mathrm{o,set}^\mathrm{dc}-\Delta v_\mathrm{o}^\mathrm{dc}\right),  \label{eq:DC_droop}
\end{equation}
where $\Delta \hat{i}_\mathrm{f}^\mathrm{dc}$ denotes the calculated inductor current setpoint passed to the current controller, $\Delta i_\mathrm{f,set}^\mathrm{dc}$ represents the desired current reference, and $K_\mathrm{d}^\mathrm{dc}$ is the droop gain. The voltage deviation between the measured DC output voltage $\Delta v_\mathrm{o}^\mathrm{dc}$ and its reference value $\Delta v_\mathrm{o,set}^\mathrm{dc}$ determines the adjustment of the current reference, thereby regulating power injection or absorption in a decentralized manner.

Within the current controller bandwidth, the inner current controller regulates the inductor current such that it tracks its reference, i.e., $\Delta i_\mathrm{f}^\mathrm{dc} = \Delta \hat{i}_\mathrm{f}^\mathrm{dc}$. Under this assumption, combining \eqref{eq:DC_droop} with the output capacitor dynamics of Fig.~\ref{fig:converters}b:
\begin{equation}
    \Delta i_\mathrm{c}^\mathrm{dc} = C_\mathrm{dc}\ddt{\Delta v_\mathrm{o}^\mathrm{dc}} = \Delta \hat{i}_\mathrm{f}^\mathrm{dc} - \Delta i_\mathrm{o}^\mathrm{dc},
    \label{eq:DC_capacitor}
\end{equation}
yields the differential equation governing the terminal voltage dynamics of the converter:
\begin{equation}
    C_\mathrm{dc} \ddt{\Delta v_\mathrm{o}^\mathrm{dc}}=\Delta i_\mathrm{f,set}^\mathrm{dc}-\Delta i_\mathrm{o}^\mathrm{dc} + K_\mathrm{d}^\mathrm{dc} \left(\Delta v_\mathrm{o,set}^\mathrm{dc}-\Delta v_\mathrm{o}^\mathrm{dc}\right). \label{eq:DC_power}
\end{equation}
This equation describes the current-regulation mechanism of the DC I--V droop converter, as illustrated by the block diagram in Fig.~\ref{fig:droop_control}b, where the green blocks correspond to the droop control defined in \eqref{eq:DC_droop}, and the gray blocks represent the capacitor-induced dynamics described by \eqref{eq:DC_capacitor}.

\section{Duality of AC Grid-Forming and DC I--V Droop Control} \label{sec:duality}
With the converter structures and control architectures elaborated in the previous sections, the foundation for establishing the duality between AC GFM and DC I--V droop is in place. In this section, the duality is further examined from multiple perspectives, including the small-signal models of the converters, their inner control loop structures, the underlying power-sharing mechanisms, and the considered external disturbances.

\subsection{Duality of Device-Level Dynamics} \label{sec:duality_device}
The small-signal model of the converters is defined in Section~\ref{sec:preliminaries_ss} by~\eqref{eq:ilf_general} and~\eqref{eq:vdc_general}, where the corresponding transfer functions are represented by gray blocks in Fig.~\ref{fig:control_loops} and defined in \eqref{eq:Gdi_Goi} and~\eqref{eq:Gdv_Zo}. Under the considered $dq$ decoupling and focusing on the $d$-axis of the AC GFM converter, these relationships and transfer functions are identical to those of the DC half-bridge converter, with differences arising only in the values of the filter inductance and capacitances. Hence, the converter dynamics of both systems are equivalent, despite their different operating domains, i.e., three-phase AC and single-phase DC.

Based on this unified small-signal model, the inner current control design can follow an identical internal model control approach. Specifically, as shown in Sections~\ref{sec:cc_AC} and~\ref{sec:cc_DC}, the current-loop gains used for tuning the PI current controllers exhibit identical transfer functions in both the detailed forms \eqref{eq:Ti_AC} and~\eqref{eq:Ti_DC} and the simplified forms \eqref{eq:Tis_AC} and~\eqref{eq:Tis_DC}, resulting in equivalent controller parameters \eqref{eq:cc_ac}~and~\eqref{eq:cc_dc}. This identical inner control loop structure is further illustrated by the small-signal control diagrams of the AC and DC converters in Fig.~\ref{fig:control_loops}.

\subsection{Duality of Power Sharing Mechanism} \label{sec:duality_psm}
The synchronization equations for AC $P$--$\,\omega$ droop control \eqref{eq:AC_Swing} and DC I--V droop control \eqref{eq:DC_power} reveal the underlying duality. In particular, the emulated frequency $\Delta \omega^\mathrm{ac}$ in AC systems corresponds to the converter output voltage $\Delta v_\mathrm{o}^\mathrm{dc}$ in DC systems, making these variables isomorphic in their respective domains. 
Moreover, the synchronization signal governing power sharing is derived from the measured active power deviation $\Delta p_\mathrm{o}^\mathrm{ac}$ in AC systems, while the measured output current deviation $\Delta i_\mathrm{o}^\mathrm{dc}$ serves as the corresponding dual quantity in the DC domain, where $\Delta i_\mathrm{o}^\mathrm{dc}$ is effectively equivalent to $\Delta p_\mathrm{o}^\mathrm{dc}$ in p.u. due to tight DC terminal voltage regulation.
Additionally, the damping term $K_\mathrm{d}^\mathrm{ac}$ in \eqref{eq:AC_Swing} equals the droop gain $K_\mathrm{d}^\mathrm{dc}$ in \eqref{eq:DC_power}, and the virtual inertia constant $H$ corresponds to the physical DC output capacitance $C_\mathrm{dc}$.

The duality holds provided that the inner control loops of both systems exhibit comparable dynamics, which is ensured by tuning the controllers via internal model control to compensate for their open-loop behavior, as detailed in Sections~\ref{sec:cc_AC} and~\ref{sec:cc_DC}. Furthermore, due to the dual-loop structure of the AC GFM converter, its overall control dynamics can be made to approximate those of the single-loop DC I--V droop by selecting a sufficiently fast voltage controller. In particular, a proper time-scale separation between the outer voltage and droop control loops must be preserved for the AC GFM converter. This is achieved by selecting the active-power low-pass filter cut-off frequency $\omega_\mathrm{c}$ to be significantly lower than the voltage controller bandwidth $\omega_{\beta_\mathrm{v}}^\mathrm{ac}$. Since $C_\mathrm{dc}$ directly determines $H$, and $\omega_\mathrm{c}$ follows from \eqref{eq:swing_parameters}, the capacitance $C_\mathrm{dc}$ must be sufficiently large to ensure decoupling between voltage control and droop dynamics in the AC system, i.e., $\omega_\mathrm{c} \ll \omega_{\beta_\mathrm{v}}^\mathrm{ac}$. Finally, all quantities in \eqref{eq:AC_Swing} and \eqref{eq:DC_power}, including setpoints $\Delta (\cdot)_\mathrm{set}$, must be consistently defined in p.u. to preserve the~duality.

In summary, the two power-sharing mechanisms are direct duals of each other: the AC mechanism tracks $\Delta \omega^\mathrm{ac}$ based on variations in $\Delta p_\mathrm{o}^\mathrm{ac}$, while the DC mechanism droops $\Delta v_\mathrm{o}^\mathrm{dc}$ in response to $\Delta i_\mathrm{o}^\mathrm{dc}$. The former is governed by the selected $H$ and $K_\mathrm{d}^\mathrm{ac}$, while the latter is determined by $C_\mathrm{dc}$ and $K_\mathrm{d}^\mathrm{dc}$. This duality is summarized by the following relation:
\begin{equation}
\left.
\begin{aligned}
    2H &= C_\mathrm{dc}  \\
    K_\mathrm{d}^\mathrm{ac} &= K_\mathrm{d}^\mathrm{dc} \\ 
    \Delta p_\mathrm{o}^\mathrm{ac} &\leftrightarrow \Delta i_\mathrm{o}^\mathrm{dc}
\end{aligned}
\right\}
\ \Delta \omega^\mathrm{ac} \xLeftrightarrow[{\mathrm{DC} \, \eqref{eq:DC_power}}]{{\mathrm{AC} \, \eqref{eq:AC_Swing}}}\Delta v_\mathrm{o}^\mathrm{dc}, \label{eq:duality}
\end{equation}
and highlighted by the corresponding block diagrams in Fig.~\ref{fig:droop_control}.

\subsection{Duality of External Disturbance} \label{sec:duality_dist}
According to the duality relationship established in \eqref{eq:duality}, the variable dual to an active power disturbance at the AC converter terminals, denoted by $\Delta p_\mathrm{o}^\mathrm{ac}$, is an output current disturbance $\Delta i_\mathrm{o}^\mathrm{dc}$ in DC systems. Based on this intuition, the AC converter is connected to a resistive load that introduces an active power disturbance at the converter terminals, as illustrated in Fig.~\ref{fig:converters}a. Similarly, the DC converter is connected to a constant current source that imposes a disturbance in the converter output current, as shown in Fig.~\ref{fig:converters}b.

\subsection{Physical Role of AC and DC Output Capacitance} \label{sec:duality_c}
Interestingly, although not immediately obvious, the output capacitances of the AC and DC converters do not need to be equal for the duality to hold, since they serve different physical roles: in the former case, the capacitance $C_\mathrm{f}$ corresponds to the output filter capacitor, whose primary function is to attenuate switching harmonics and shape the AC terminal voltage dynamics, while $C_\mathrm{dc}$ directly impacts energy storage and terminal voltage dynamics, and thus plays a role analogous to the inertia constant of rotating machines. The influence of $C_\mathrm{f}$ on overall system dynamics is largely mitigated by the AC GFM voltage controller due to internal model control, while $C_\mathrm{dc}$ determines the inertia constant $H$ via \eqref{eq:duality}.

\section{Simulation-Based Verifications} \label{sec:simulations}
To verify the theoretical analysis of the AC–DC duality and the associated controller-tuning methodology, time-domain simulations are conducted in \textsc{Matlab}/\textsc{Simulink}. The simulations are based on voltage-averaged models of the AC and DC converters shown in Fig.~\ref{fig:converters}, where the power electronic interfaces (i.e., DC half-bridge and AC three-phase inverter) are represented using ideal switching functions, thereby neglecting switching dynamics. The effect of the PWM is approximated by a first-order delay with a time constant equal to one sampling period. The sampling frequency is set to $\SI{100}{k Hz}$, and simulations are performed with a fixed time step of $\SI{1}{\micro s}$.

\subsection{System Setup}
Both systems operate at an output power of $\SI{2000}{W}$ with an input voltage of $\SI{700}{V}$ and an output voltage of $\SI{350}{V}$ (DC for the DC system and RMS line-to-line for the AC system), while the AC system operates at a nominal frequency of $50~\mathrm{Hz}$. The base values are selected as $\SI{4000}{W}$ and $\SI{350}{V}$, yielding a base current of $\SI{11.4286}{A}$. Both converters employ an output filter inductance of $\SI{7.7}{mH}$, with capacitance $C_\mathrm{f}=\SI{0.72}{m F}$ for the AC system and $C_\mathrm{dc}=\SI{72}{mF}$ for the DC system. As discussed in Section~\ref{sec:duality_c}, the capacitances are not required to be equal due to their distinct physical roles.

To ensure comparable closed-loop behavior between the AC and DC systems, the current-loop bandwidth of both converters is selected as $1\%$ of the switching frequency, i.e., $\omega_{\beta_\mathrm{i}}^\mathrm{ac} = \omega_{\beta_\mathrm{i}}^\mathrm{dc} = 0.01 \cdot 2\pi f_\mathrm{s}$, where $f_\mathrm{s}=\SI{50}{kHz}$ denotes the switching frequency assumed for current controller tuning. The integral time-constant tuning factors $c_\mathrm{i}$ are set to 20 for both converters. Moreover, the voltage-loop bandwidth of the AC GFM converter is chosen as a fixed fraction of the current-loop bandwidth, $\omega_{\beta_\mathrm{v}}^\mathrm{ac} = 0.2\,\omega_{\beta_\mathrm{i}}^\mathrm{ac}$, and the corresponding tuning factor $c_\mathrm{v}^\mathrm{ac}$ is set to 2.5. Finally, the DC I--V droop gain is set to 0.75, while the AC $P$--$\,\omega$ droop gain follows from \eqref{eq:duality} and \eqref{eq:swing_parameters}. For the selected $C_\mathrm{dc}$, the duality relation \eqref{eq:duality} yields $H=\SI{1.1025}{s}$, and the low-pass filter cut-off frequency follows from \eqref{eq:swing_parameters}.

%%%%%%%%%%%%%%%%%%%%%%%%%%%%%%%%%%%%%%%%%%%%%%%%%%%%%%%%%%%%%%%%%
\begin{figure}[b!]
    \centering
    \vspace{-0.10cm}
    \includegraphics[width=0.485\textwidth]{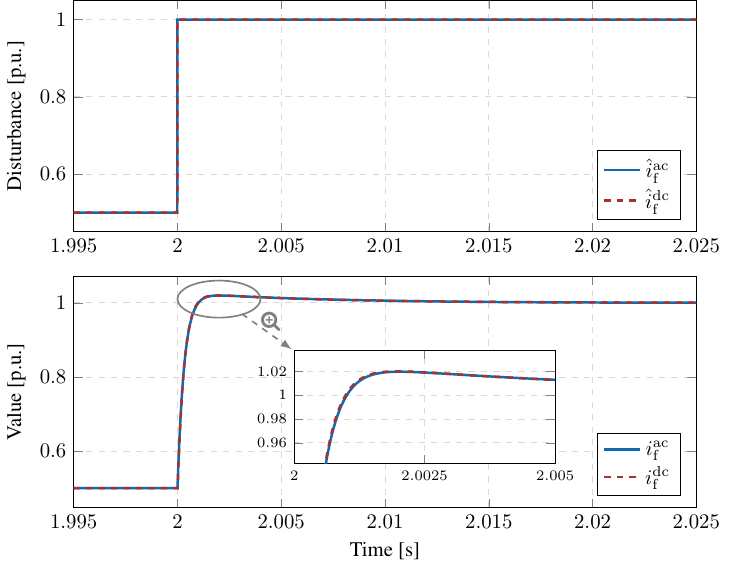}
    \vspace{-0.70cm}
    \caption{Current controller response of the AC and DC converters.}
    \label{fig:cc_response}
\end{figure}
%%%%%%%%%%%%%%%%%%%%%%%%%%%%%%%%%%%%%%%%%%%%%%%%%%%%%%%%%%%%%%%%%

\subsection{Current Controller Response}
The equivalence of the inner current controller tuning is verified by applying a step change to the inductor current reference at $t=\SI{2}{s}$ and comparing the resulting current responses of the AC and DC converters, namely the $d$-axis AC inductor current $i_{\mathrm{f}}^\mathrm{ac}$ and the DC inductor current $i_{\mathrm{f}}^\mathrm{dc}$.

The applied reference step and corresponding current responses are shown in Fig.~\ref{fig:cc_response}. The nearly identical trajectories, both during transients and in steady state, demonstrate the duality between the AC and DC current controllers under the internal model control tuning of Sections~\ref{sec:cc_AC} and~\ref{sec:cc_DC}.

%%%%%%%%%%%%%%%%%%%%%%%%%%%%%%%%%%%%%%%%%%%%%%%%%%%%%%%%%%%%%%%%%
\begin{figure}[t!]
    \centering
    \includegraphics[width=0.485\textwidth]{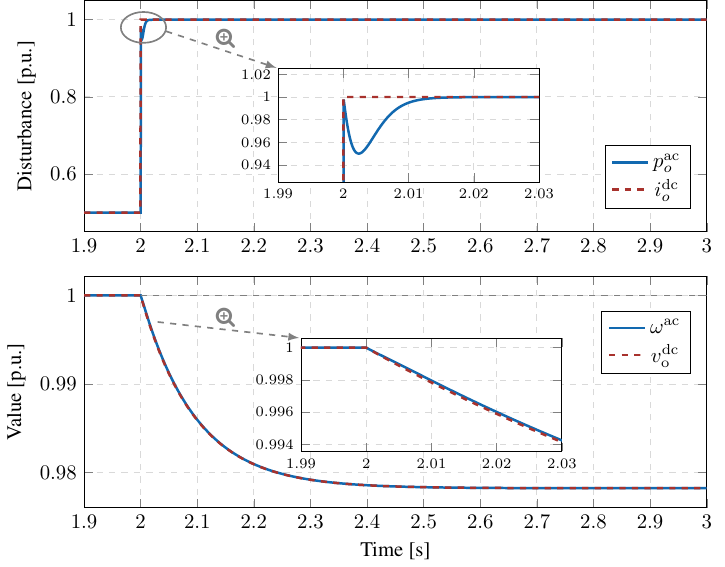}
    \vspace{-0.60cm}
    \caption{Disturbance response of the AC and DC converters.}
    \label{fig:dynamic_response}
    \vspace{-0.10cm}
\end{figure}
%%%%%%%%%%%%%%%%%%%%%%%%%%%%%%%%%%%%%%%%%%%%%%%%%%%%%%%%%%%%%%%%%

%%%%%%%%%%%%%%%%%%%%%%%%%%%%%%%%%%%%%%%%%%%%%%%%%%%%%%%%%%%%%%%%%%%%%%%%%%%
\begin{figure}[t!]
    \centering
    \includegraphics[width=0.485\textwidth]{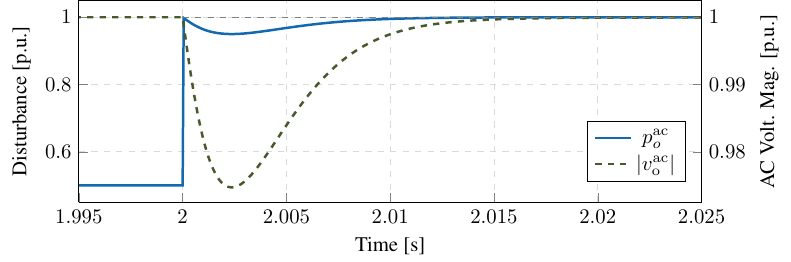}
    \vspace{-0.70cm}
    \caption{Voltage response of the AC converter under disturbance.}
    \label{fig:dynamic_response_vAC}
    \vspace{-0.20cm}
\end{figure}
%%%%%%%%%%%%%%%%%%%%%%%%%%%%%%%%%%%%%%%%%%%%%%%%%%%%%%%%%%%%%%%%%

\subsection{Dynamic Response}
Finally, the equivalence between the AC frequency and DC voltage dynamics is verified by subjecting both converter systems to equivalent external disturbances, as discussed in Section~\ref{sec:duality_dist}. Specifically, at $t=\SI{2}{s}$, a step increase in active power $p_\mathrm{o}^\mathrm{ac}$ is applied to the AC converter, while the DC converter is subjected to an increase in output current $i_\mathrm{o}^\mathrm{dc}$.

The upper plot of Fig.~\ref{fig:dynamic_response} illustrates the applied disturbances, while the lower plot shows the corresponding responses of the AC system frequency $\omega^\mathrm{ac}$ and the DC bus voltage $v_\mathrm{o}^\mathrm{dc}$. As observed, the frequency and voltage trajectories exhibit nearly identical transient and steady-state behavior.

Moreover, as highlighted in the zoomed-in view of the upper plot in Fig.~\ref{fig:dynamic_response}, the active power disturbance in the AC case is not applied instantaneously. Specifically, when the resistive load is increased, the terminal voltage $|v_\mathrm{o}^\mathrm{ac}|$ of the GFM converter initially decreases due to the limited bandwidth of the voltage controller, which is unable to maintain $|v_\mathrm{o}^\mathrm{ac}|$ at its setpoint value of $1$~p.u. during the disturbance, as indicated by the dashed line in Fig.~\ref{fig:dynamic_response_vAC}. This transient voltage drop causes a temporary deviation in the converter output power, which is restored to its setpoint once the $q$-axis reactive power control loop regulates the voltage magnitude back to its nominal value. In contrast, the disturbance in the DC system is applied via an ideal current source, resulting in an immediate change in $i_\mathrm{o}^\mathrm{dc}$, and thus, $v_\mathrm{o}^\mathrm{dc}$.

\section{Conclusion} \label{sec:concl}
This paper presents a novel perspective on the duality between AC GFM control and DC I--V droop control in converter-based microgrids. We show that voltage and frequency act as dual variables, while current and active power serve as the common synchronization signal in both domains. Theoretical analysis and simulation results confirm the structural similarities and demonstrate how tuning principles can be aligned under a unified framework. These findings provide a foundation for harmonized control strategies, improved interoperability, and stability in future hybrid AC/DC microgrids.

% References section
\bibliographystyle{IEEEtran}
\bibliography{bibliography}

% That's all folks
\end{document}